\documentclass[review,authoryear]{elsarticle}

\usepackage{graphicx}
\usepackage{caption}
\usepackage{subcaption}
\usepackage{siunitx}
\usepackage{hyperref}

\usepackage[export]{adjustbox}

\begin{document}

\begin{frontmatter}

\title{Luminous Efficiency Estimates of Meteors -I. Uncertainty analysis}


\author[mymainaddress]{Dilini Subasinghe}\corref{mycorrespondingauthor}
\cortext[mycorrespondingauthor]{Corresponding author}
\ead{dsubasi@uwo.ca}

\author[mymainaddress,mysecondaryaddress]{Margaret Campbell-Brown}

\author[mymainaddress,mysecondaryaddress]{Edward Stokan}


\address[mymainaddress]{Department of Physics and Astronomy, University of Western Ontario, London, Canada, N6A 3K7}
\address[mysecondaryaddress]{Centre for Planetary Science and Exploration, University of Western Ontario, London, Canada, N6A 5B7} 

\begin{abstract}
The luminous efficiency of meteors is poorly known, but critical for determining the meteoroid mass. We present an uncertainty analysis of the luminous efficiency as determined by the classical ablation equations, and suggest a possible method for determining the luminous efficiency of real meteor events. We find that a two-term exponential fit to simulated lag data is able to reproduce simulated luminous efficiencies reasonably well. 
\end{abstract}

\begin{keyword}
\texttt meteors meteoroids optical masses 
\MSC[2010] 00-01\sep  99-00
\end{keyword}

\end{frontmatter}


\section{Introduction}
Determining the mass of a meteoroid, a basic property, is currently very difficult to do. Because most meteoroids are too small to reach the ground, meteoroid mass needs to be determined through observations. The simplest method is to use the total luminous energy emitted during ablation. The large uncertainty associated with mass is due to many unknown variables, such as the bulk density, shape, and luminous efficiency, and their (possible) changes during ablation. Spacecraft hazard estimates rely on accurate meteoroid masses: while rare, collisions and damage to satellites by meteoroids have occurred \citep{Caswell1995}. \newline

There are two coupled differential equations in classical meteor physics that describe the state of the meteoroid and allow mass to be determined: the luminous intensity equation and the drag equation. The luminous intensity, given in Equation \ref{eq:luminous}, assumes the brightness (or luminous intensity, $I$) of a meteor is proportional to the change in kinetic energy.  \newline

\begin{equation}
I =-\tau\frac{{\rm d}E_{k}}{{\rm d}t} = -\tau\left(\frac{v^{2}}{2}\frac{{\rm d}m}{{\rm d}t} + mv\frac{{\rm d}v}{{\rm d}t}\right)
\label{eq:luminous}
\end{equation} 

The proportionality constant, $\tau$, is the luminous efficiency, the fraction of kinetic energy dissipated as meteor light. The $m$ refers to the total meteoroid mass, including any fragments. Despite their small masses ($< 10^{-4}$ kg), the majority of small meteoroids do fragment \citep{Subasinghe2016}, and that light is taken into account when calculating the \textit{photometric mass}. \newline

Equation \ref{eq:luminous} may be rearranged to solve for the photometric mass, but there is typically a large associated uncertainty, due to the vast range in luminous efficiency values. The second term in Equation \ref{eq:luminous} is often neglected, as the deceleration for fast, faint meteors is negligible, relative to the first term. Using typical values, it can be shown that for slow meteors the deceleration term is almost equal in importance to the mass loss term, but becomes significantly less important at higher speeds (i.e. the deceleration term is about 40\% of the mass loss term for a meteor moving at 11 km/s, but only 1\% for a meteor travelling at 70 km/s). \newline

The drag equation given in Equation \ref{eq:drag}, can also be used to determine the mass of a meteoroid, and is derived through conservation of momentum. \newline

\begin{equation}
\label{eq:drag}
\frac{\mathrm{d}v}{\mathrm{d}t} = -\frac{\Gamma \rho_{atm} v^2 A}{m^\frac{1}{3}{\rho_m^\frac{2}{3}}}
\end{equation}

The mass in this equation is called the \textit{dynamic mass}, as it is based on the deceleration of the largest, brightest fragment (or group of similarly sized fragments). The other variables in Equation \ref{eq:drag} are the drag coefficient $\Gamma$, the atmospheric density $\rho_{atm}$, the velocity $v$, the shape factor $A$, and the meteoroid density $\rho_{m}$. Previous studies have found that the dynamic mass of faint meteors is consistently smaller than the photometric mass, and is thus not an accurate measure of the true meteoroid mass for fragmenting meteors \citep{Verniani1965}. Again, this is because the photometric mass considers the mass of all light producing fragments, and the dynamic mass only considers the largest, brightest fragment.\newline

Since most meteoroids do fragment, it is therefore useful to better understand the luminous efficiency to determine the meteoroid mass through the luminous intensity equation. The goal of this study is ultimately to examine faint meteoroids that do not appear to fragment, to determine their luminous efficiencies. In those cases, the dynamic mass, found by the deceleration, is equivalent to the photometric mass, and we can solve for the luminous efficiency. This luminous efficiency can then be used to find the masses of other meteoroids, even those that fragment. It has been suggested that the luminous efficiency depends on meteoroid speed and height, camera spectral response (an iron-rich meteoroid may emit strongly in the blue portion of the visible spectrum, but may not be detected if the camera system is not sensitive to that range), meteoroid and atmospheric composition, and possibly meteoroid mass, among other factors, but the extent to which it depends on each variable is unknown \citep{Ceplecha1998}. \newline

\subsection{Previous luminous efficiency studies}
As a meteoroid enters the atmosphere, it heats up through collisions with atmospheric atoms and molecules. This results in meteoroid ablation and the release of meteoric atoms and molecules into the atmosphere. Evaporated meteoritic material interacts with atmospheric molecules or other ablated atoms, leading to the excitation of the meteoric and atmospheric species. The luminosity observed is due to the decay of these excited states and is emitted in spectral lines. \newline

\begin{figure}
\caption{Some of the past work done on luminous efficiency using various methods (lab experiments; artificial meteors; radar and optical observations). Note that the luminous efficiency values (given as a percentage) are shown on a log scale: the large discrepancies between luminous efficiency values for a given meteor speed cause large uncertainties in the derived mass. The constant 0.7\% luminous efficiency corresponds to the value used in this study for the simulated meteor events.}
\includegraphics[width=\textwidth]{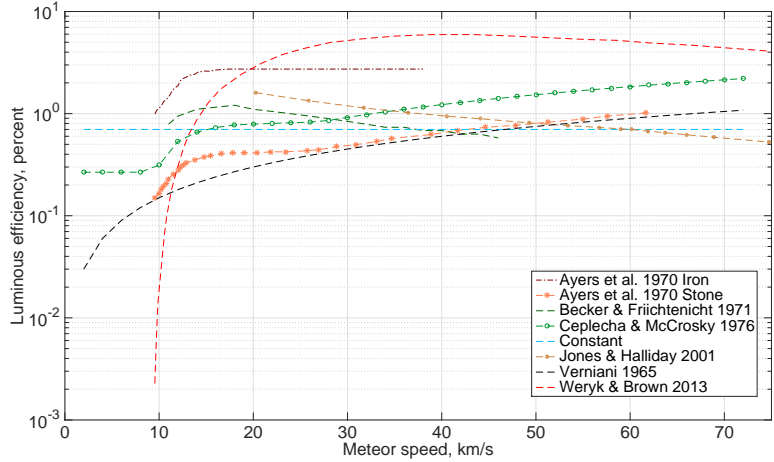}
\label{fig:historical_plot}
\end{figure}

Many of the early luminous efficiency studies were done by Opik, who used a theoretical approach to determine luminous efficiencies for various atoms. Uncertainty in the approach used led to questions of the validity of his work: he is mentioned here for completeness. \citet{Verniani1965} combined the drag equation with the luminous intensity equation to solve for the luminous efficiency. This method explicitly equates the photometric mass with the dynamic mass, which is problematic since these masses are not equivalent for meteoroids that fragment, and studies have shown that the majority of observed meteoroids show fragmentation \citep{Subasinghe2016, Weryk2013}. \citet{Verniani1965} attempted to correct for fragmentation, and assumed that luminous efficiency can be described as shown in Equation  \ref{eq:velocitytau}, with luminous efficiency proportional to speed raised to some power. He found for the 413 Super Schmidt meteors he studied, that $n=1.01 \pm 0.15$ and $1.24 \pm 0.22$ for fragmenting and non-fragmenting meteors respectively. He further investigated whether luminous efficiency depends on mass (he found that it does not), and confirmed that luminous efficiency does not depend on the atmospheric density. He used a single non-fragmenting meteor, suggested to be asteroidal in origin (based on orbital characteristics), to conclude that in the photographic bandpass, the constant $\tau_{0}$ in Equation \ref{eq:velocitytau}, is $\log_{10} \tau_{0} = -4.37 \pm 0.08$ for n = 1. These results, along with the following studies, are illustrated in Figure \ref{fig:historical_plot}. \newline

\begin{equation}
\tau = \tau_{0}v^{n}
\label{eq:velocitytau}
\end{equation}

Many lab experiments were performed in the sixties and seventies, with the obvious advantage of being able to control many aspects of the ablation process such as the mass and composition of the ablating particles, and the gas density in which the particles ablate. One of the limitations of lab experiments for luminous efficiency estimates is the difficulty in reaching all valid meteor speeds -- \citet{Friichtenicht1968} reached speeds between 15 - 40 km/s, while \citet{Becker1971} explored speeds between 11 - 47 km/s. The experimental lab set up involved charging and accelerating particles in a Van deGraaf generator (detectors measured the charge and velocity), and then observing as the particles ablated in a gas region meant to simulate free molecular flow (13.3 Pa). \citet{Becker1971} used 167 iron and 120 copper spherical simulated meteors, with diameters between 0.05 and 1 micron, and their results are shown in Figure \ref{fig:historical_plot}. \citet{Becker1973} used essentially the same methods as \citet{Becker1971}, but studied silicon and aluminium particles with similar diameters, as they ablated in a gas region of air, nitrogen, or oxygen, at a pressure around 27 Pa. These results are not applicable to optical meteors directly, as these lab studies used particles much smaller than the millimetre sized objects that most optical cameras observe, and the pressures at which the micron sized particles ablated correspond to heights much lower (between 55 - 65 km) than those at which optical cameras typically observe (around 90 - 110 km). \newline

Artificial meteors are another method of determining the luminous efficiency. In this method, objects of known mass and composition are subjected to atmospheric re-entry, and observed as they ablate. \citet{Ayers1970} used iron and nickel objects, launched between 1962 and 1967, observing a total of ten artificial meteors. These artificial meteoroids had either a disk or cone shape, and their masses ranged between 0.64 - 5.66 grams. The average begin and end heights were 76 and 66 km, respectively. These artificial meteoroids were observed optically, and the luminous intensity and velocity were collected. Combined with the measured initial meteoroid mass, the luminous efficiency was calculated using a simplified version of Equation \ref{eq:luminous}, in which the second term (related to the deceleration) is ignored. \citet{Ayers1970} found that $n=1.9 \pm 0.4$ in Equation \ref{eq:velocitytau} for four artificial meteors, including one from \citet{McCrosky1963}. \citet{Ayers1970} also formulated a luminous efficiency relationship for meteors of stony composition, assuming that 15\% of the mass is iron, which is the main emitter in their blue sensitive cameras: that between 20 and 30 km/s, the luminous efficiency increases monotonically; and that above 30 km/s, $n=1$. This may not be applicable to other more red-sensitive optical systems. They noted that this work was a first approximation. A slight reworking of the \cite{Ayers1970} results was done by \citet{Ceplecha1976}, who increased the proportion of iron by weight from 15\% to 28\%. The luminous efficiency suggested by \citet{Ceplecha1976} is a piece-wise function (shown in Figure \ref{fig:historical_plot}), and was used for fireball analysis. \newline

\citet{Jones2001} defined an excitation coefficient, which is the average number of times a meteoric atom is excited during ablation. In combining theory and lab measurements, they found that their primary excitation probability is unphysical beyond 42 km/s (they assumed ionised atoms are unavailable for excitation). They referred to scattering and diffusion cross-sections to describe the excitation coefficient, but found that the values were higher than experimental values suggested. \newline

Simultaneous optical and radar observations of meteors were used by \citet{Weryk2013a} to determine luminous efficiency for the bandpass of their GEN-III image intensifiers. The ratio of the ionisation coefficient $\beta$ (the number of electrons produced per ablating atom) to the luminous efficiency $\tau$ can be determined through radar and video measurements, and assuming a value for either $\beta$ or $\tau$ allows the other to be determined. \citet{Jones1997} determined an expression for $\beta$ using both theory and observations, which \citet{Weryk2013a} used to determine a peak bolometric value of $\tau = 5.9\%$ at 41 km/s, for their Gen-III bandpass (470 - 850 nm). \newline

\section{Method}
The purpose of this work is to develop a method, using simulated data, to calculate luminous efficiency from non-fragmenting meteors observed with a high-resolution optical system, and to investigate the sensitivity of the method to the various assumed parameters. Equating the dynamic and photometric masses is appropriate, provided the meteoroid does not fragment, and allows for the determination of the luminous efficiency. The classical meteor ablation equations apply to a solid, single, non-fragmenting body. The Canadian Automated Meteor Observatory (CAMO; discussed below) has at best, a resolution of 3 m/pixel in its narrow-field optical camera, which means it can confirm that the meteor events collected do not significantly fragment on that scale. The dynamic mass can then be equated to the photometric mass to solve for the luminous efficiency: rearranging Equations \ref{eq:luminous} and \ref{eq:drag} gives us:

\begin{equation}
m= - \frac{\Gamma^3 \rho_{atm}^3 v^6 A^3}{\rho_{m}^2 (\frac{dv}{dt})^3}
\label{eq:mass}
\end{equation}

\begin{equation}
\tau = - \frac{I}{\frac{v^{2}}{2}\frac{{\rm d}m}{{\rm d}t} + mv\frac{{\rm d}v}{{\rm d}t}}
\label{eq:tau}
\end{equation}

Assumptions must be made for certain parameters: the drag coefficient $\Gamma$, which can range from 0 - 2; the shape factor $A$, given by $\frac{surface\; area}{volume^\frac{2}{3}}$; and the meteoroid density $\rho_{m}$, which can range from 1000 - 8000 kg/m$^{3}$. For the drag coefficient and the shape factor, typical values were used ($\Gamma$ = 1; $A = 1.21$ (sphere)). An atmospheric density profile was taken from the NRLMSIS E-00 model \citep{Picone2002}. \newline

\subsection{Future application to real data}
The Canadian Automated Meteor Observatory is a two station, image intensified video system, located in Ontario, Canada \citep{Weryk2013}. The two stations are approximately 45 km apart, with one station in Tavistock, Ontario, Canada (43.265$^\circ$N, 80.772$^\circ$W), and the other in Elginfield, Ontario, Canada (43.193$^\circ$N, 81.316$^\circ$W). Sky conditions permitting, the camera systems run each night. The guided system, used for data collection, consists of two cameras: a wide-field camera, with a field of view of 28$^\circ$, and a narrow-field camera, with a field of view of 1$^\circ$.5. The wide-field cameras, which run at 80 frames per second, allow for orbit determination, as well as light curve measurements; and the narrow-field cameras, which run at 110 frames per second, provide high-resolution observations of the meteoroid. To reduce the possibility of image saturation, the cameras each have 12 bit image depth. \newline

Meteors are detected in the wide-field camera in real time with the All Sky and Guided Automatic Realtime Detection (ASGARD) software\citep{Weryk2008}, and ASGARD directs a pair of mirrors to track the meteor and direct the image into the narrow-field camera. \newline

With the high-resolution narrow-field cameras, meteors that appear to show single body ablation can be selected and studied to determine their luminous efficiencies. In a future work, we will analyze a number of events and apply this luminous efficiency determination method to them. The meteor events will be reduced using mirfit: software designed to process meteor events recorded with the CAMO tracking system, and provide high-precision position measurements (sub-metre scale). \newline

\section{Sensitivity Analysis}
One of the main difficulties in solving for luminous efficiency is determining the measured deceleration of the meteor, needed for both the dynamic mass (Equation \ref{eq:mass}) and the luminous efficiency (Equation \ref{eq:tau}). Small uncertainties in the measured position result in large point-to-point errors in the speed, and very large scatter in the deceleration. 
To test the sensitivity of our technique to the assumptions made and the fitting techniques used, we simulated meteors using the model of \cite{CampbellBrown2004}. We used the classical ablation model to investigate different smoothing and fitting algorithms. The lag is the distance that the meteoroid falls behind an object with a constant velocity (equal to the initial meteoroid velocity, which is determined by fitting the first half of the distance-time data), and  requires a monotonically increasing form. As a first attempt, we expect an exponential relationship between the meteoroid lag and time, based on the atmospheric density encountered by the meteoroid increasingly roughly exponentially with time. A two term exponential will provide a better fit than a single term exponential (more terms and/or higher order terms will fit the data better, but it is important to note that adding more terms will eventually overfit the data and does not have any physical justification). \newline

\begin{figure}
    \centering
    \begin{subfigure}[b]{\textwidth}
        \includegraphics[width=\textwidth]{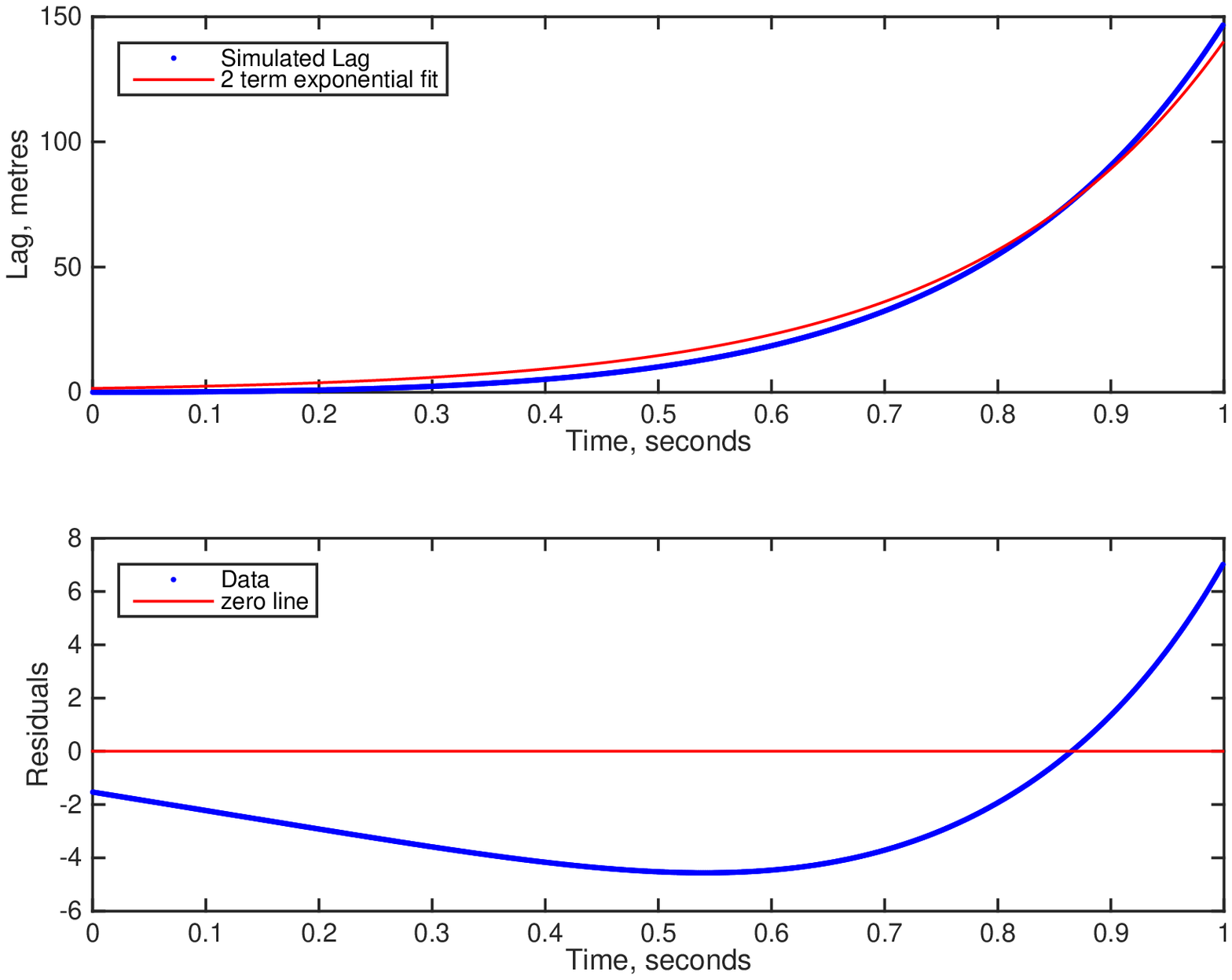}
        \caption{Fitting the entire lag with a two-term exponential.}
    \end{subfigure}
    \begin{subfigure}[b]{\textwidth}
        \includegraphics[width=\textwidth]{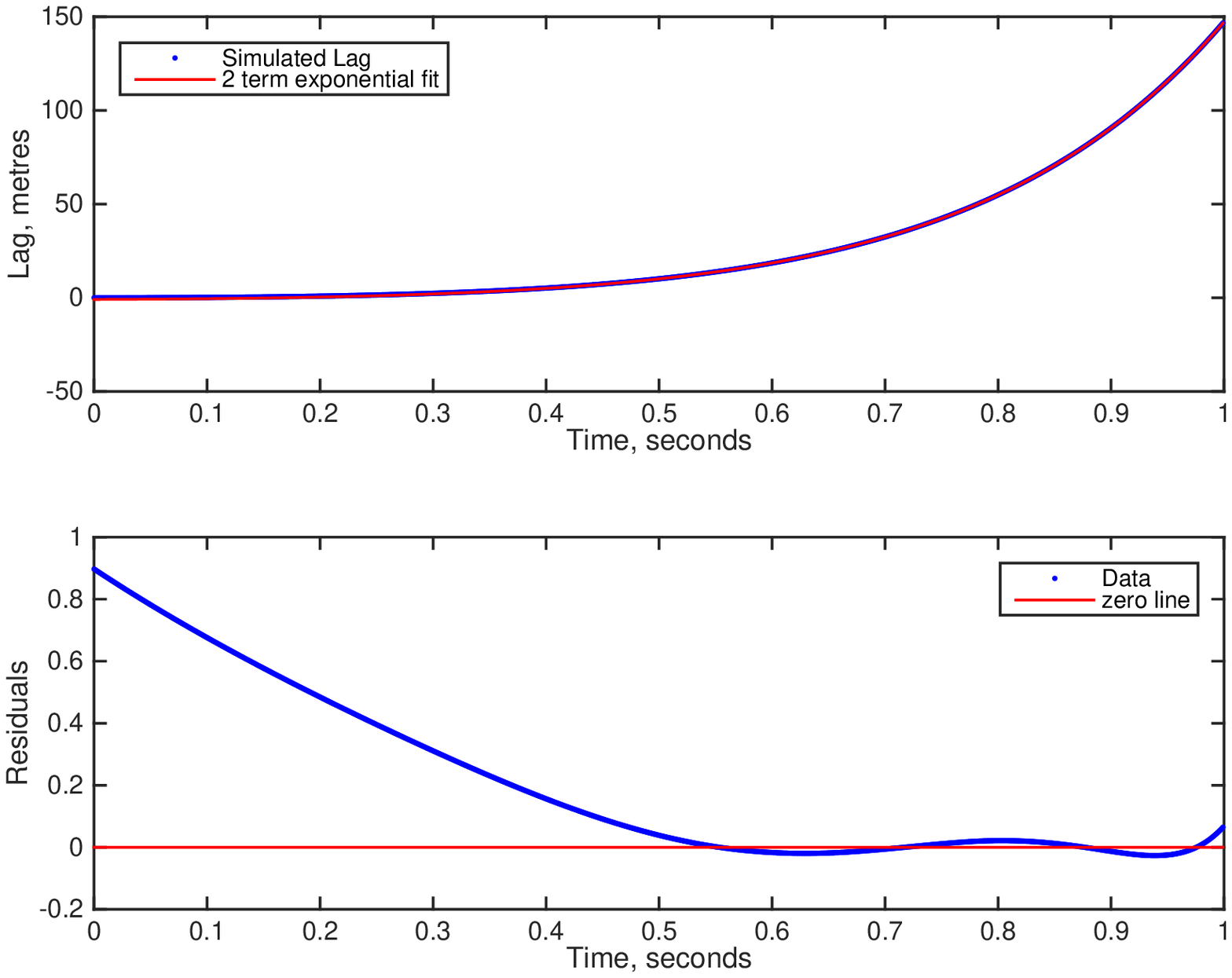}
        \caption{Fitting only the second half of the meteoroid lag, but applying the fit to the entire ablation time period for comparison.}
    \end{subfigure}
    \caption{Residual plots for fitting the entire lag and the second half of the lag, with a two-term exponential.}
\label{fig:residuals}
\end{figure}

A classically modelled meteor with the following parameters was investigated for fitting: mass of $10^{-5}$ kg, density of $2000$kg/m$^{3}$, and initial speed of $30$km/s. Because meteors show very little deceleration at the beginning of their ablation, a comparison of fitting the lag from the full curve versus the second half of the lag was done and the results are shown in Figure \ref{fig:residuals}. \newline

\begin{figure}
    \centering
    \begin{subfigure}[b]{0.70\textwidth}
        \includegraphics[width=\textwidth]{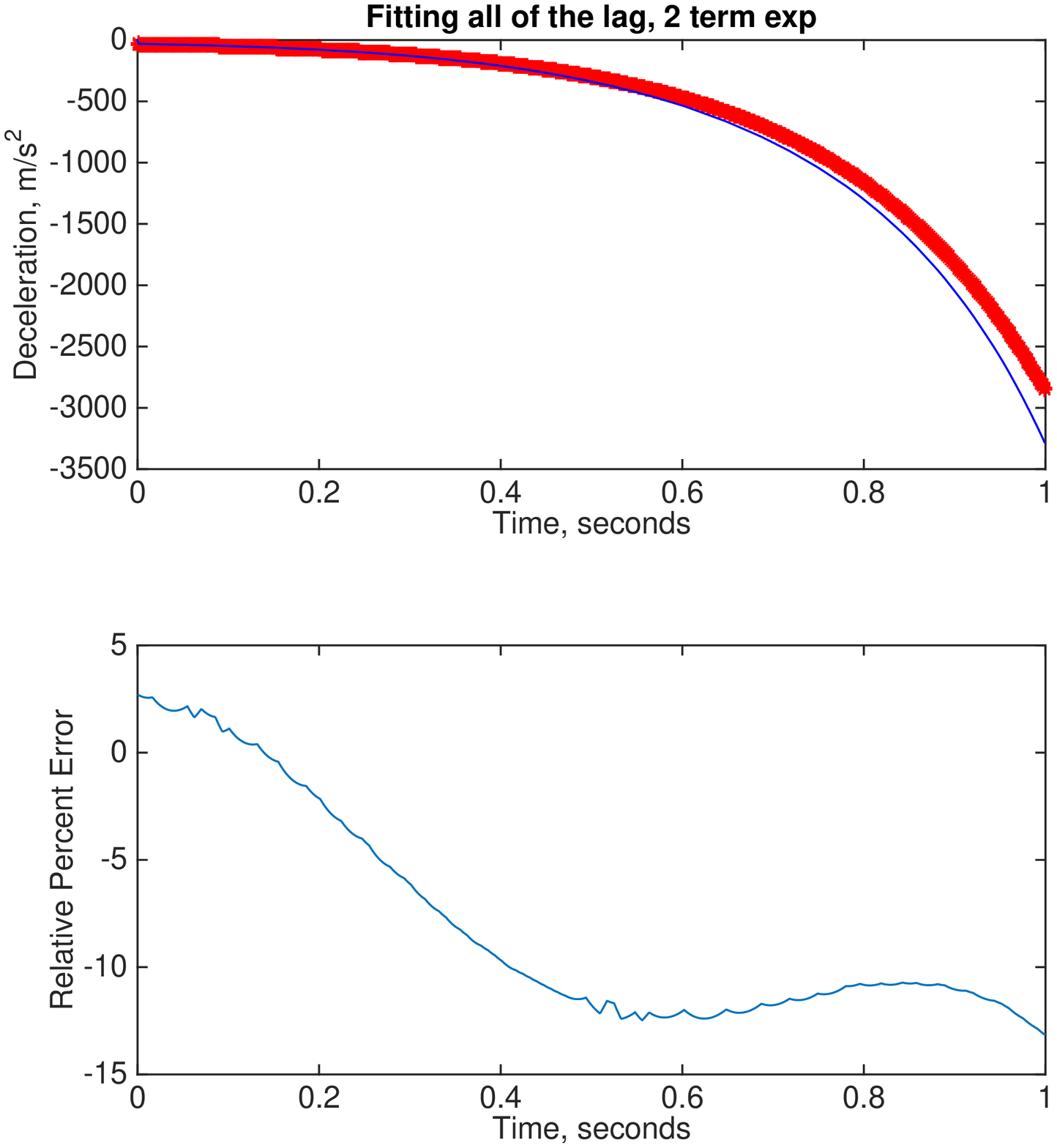}
        \caption{}
        \label{fig:full_two_term_exp}
    \end{subfigure}

    \begin{subfigure}[b]{0.70\textwidth}
        \includegraphics[width=\textwidth]{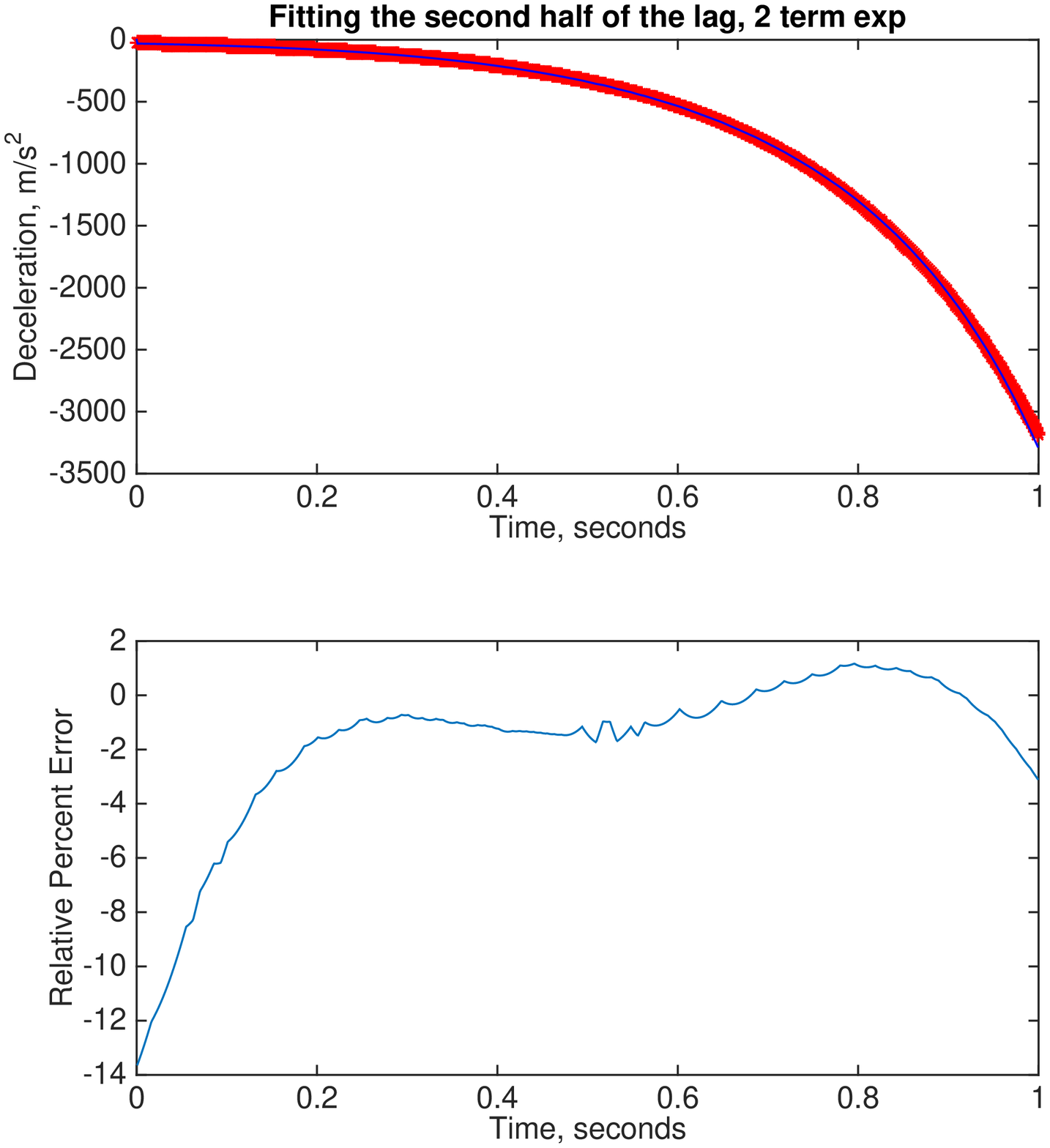}
        \caption{}
        \label{fig:half_two_term_exp}
    \end{subfigure}
    \caption{A comparison of the relative percentage error in deceleration, when fitting a two term exponential to the entire modelled meteoroid lag, versus fitting only the second half of the lag. The red points correspond to the fitted points, and the thin blue solid line shows the simulated deceleration.}
\label{fig:rel_percent_error}
\end{figure}

Fitting only the second half of the lag curve gives smaller residuals (relative to the model lag), and is more accurate at later times, when the meteoroid deceleration is more apparent and easier to fit. The RMSE value for fitting the entire lag profile was 3.49, and for fitting only the second half of the lag profile was 0.018. Because meteor ablation can last from less than a second to a few seconds, the decision was made to fit the second half of the ablation profile, rather than the last second, or half second. A comparison of the derived deceleration (based on the second derivative of the two-term exponential fit to the lag) to the model deceleration was also done, and is shown in Figure \ref{fig:rel_percent_error}. When fitting the entire lag profile, the derived deceleration matches the simulated deceleration well towards the beginning of the ablation profile, but the absolute relative error is large towards the end where deceleration is greatest, and which is of greatest interest for finding luminous efficiency. In Figure \ref{fig:half_two_term_exp}, only the second half of the lag data was fit, but the fit was extended backwards for comparison purposes. The relative percentage error is smaller when the deceleration is greatest, compared to when the entire lag is fit, as shown in Figure \ref{fig:full_two_term_exp}.\newline

Based on Figures \ref{fig:residuals} and \ref{fig:rel_percent_error}, a two-term exponential fit \(y = ae^{bx} + ce^{dx}\) to the second half of the lag data is able to visually reproduce a classically modelled meteoroid reasonably well. \newline

To investigate this method for other parameters, a set of simulated meteors were created, each with different parameters (speed, mass, shape factor, meteoroid density, drag coefficient) and tested to see if the luminous efficiency used to simulate the meteor could be extracted from simulated observations with this method. The simulated meteors were generated with the ablation model of \citet{CampbellBrown2004}. Calculation of the luminous efficiency was done blind, with no knowledge of the value used in the simulation until the analysis was complete. \newline

\begin{figure}
\centering
\begin{subfigure}{0.5\textwidth}
\centering
\caption{}
\includegraphics[width=\textwidth]{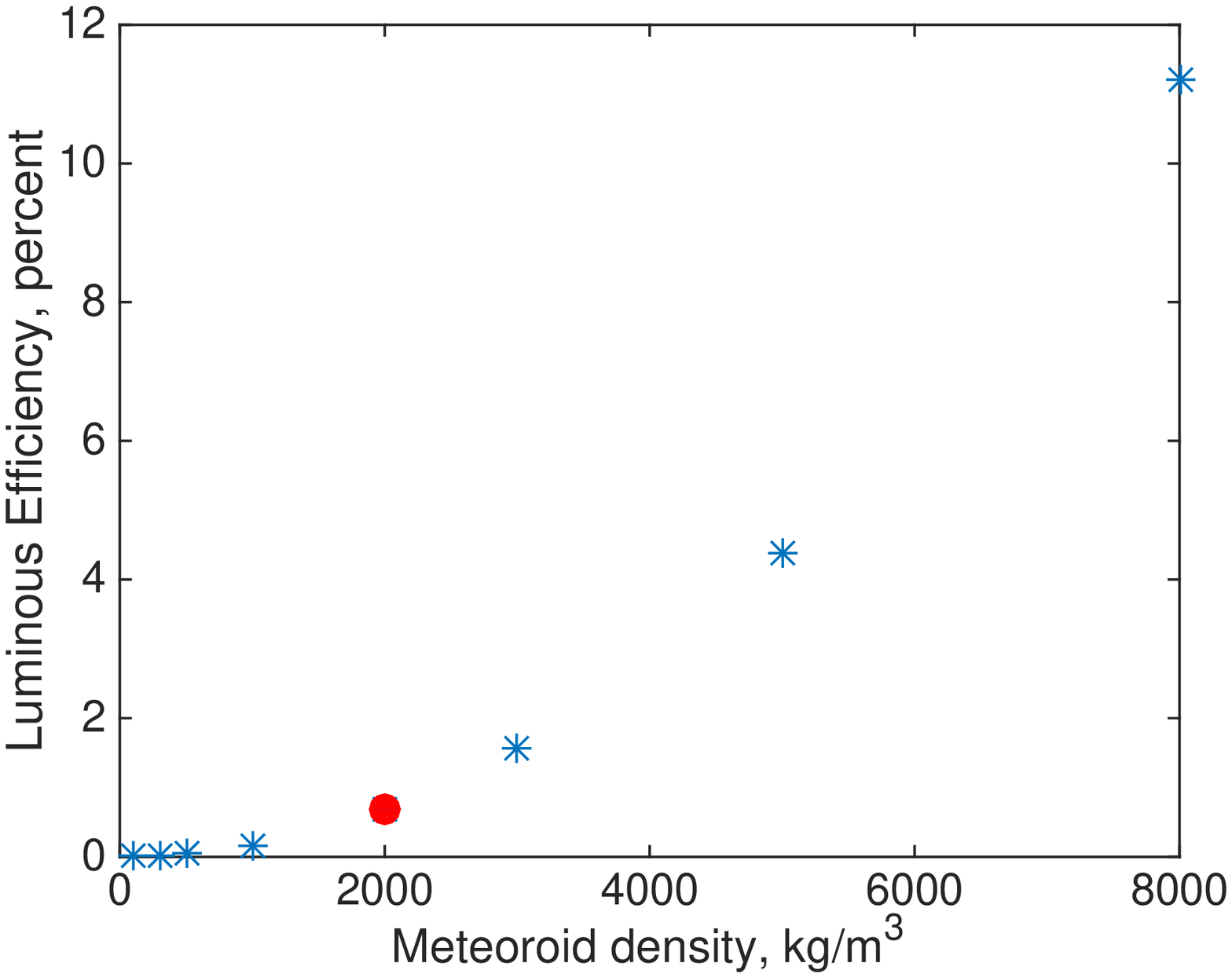}
\end{subfigure}

\begin{subfigure}{0.5\textwidth}
\centering
\caption{}
\includegraphics[width=\textwidth]{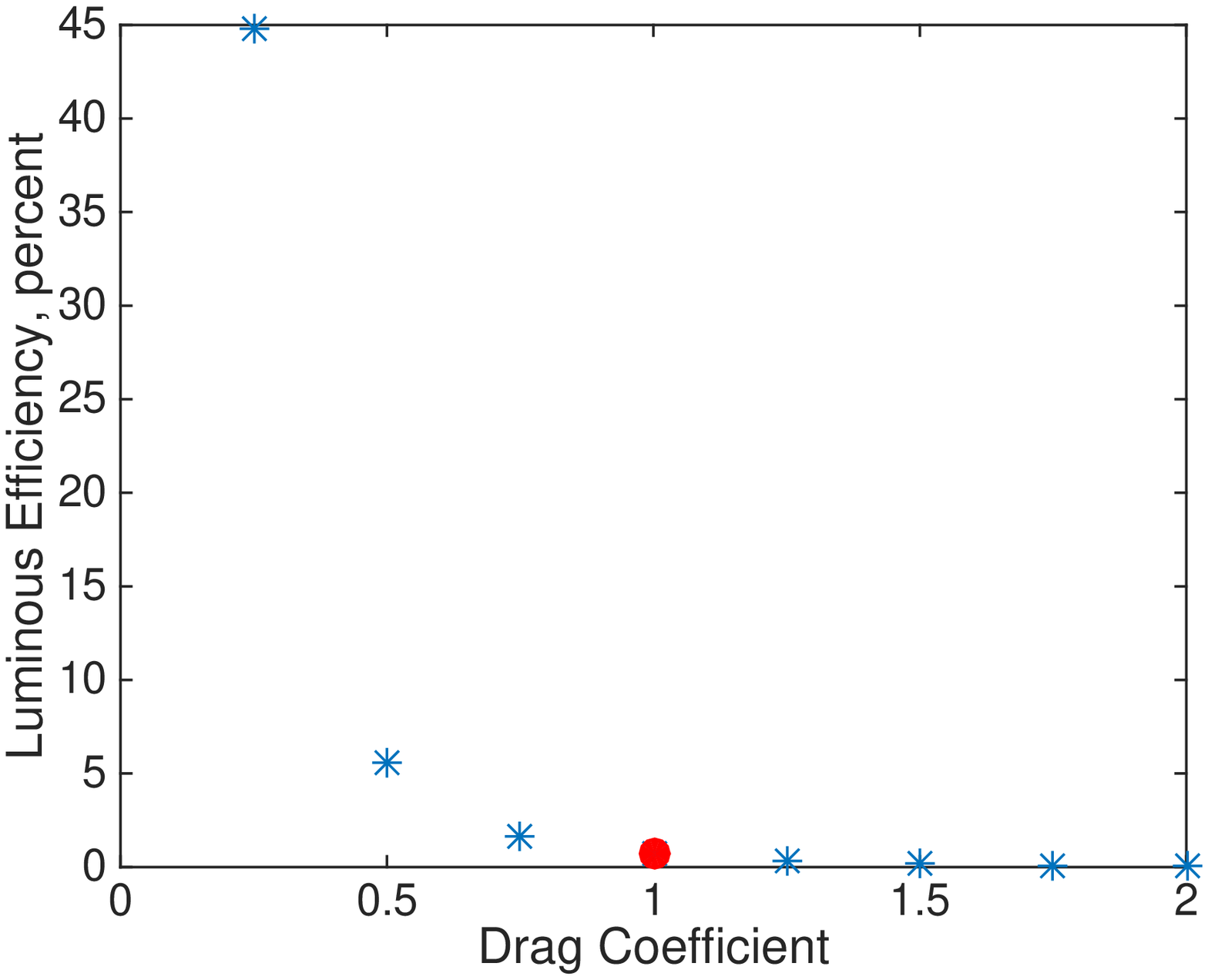}
\label{fig:drag}
\end{subfigure}

\begin{subfigure}{0.5\textwidth}
\centering
\caption{}
\includegraphics[width=\textwidth]{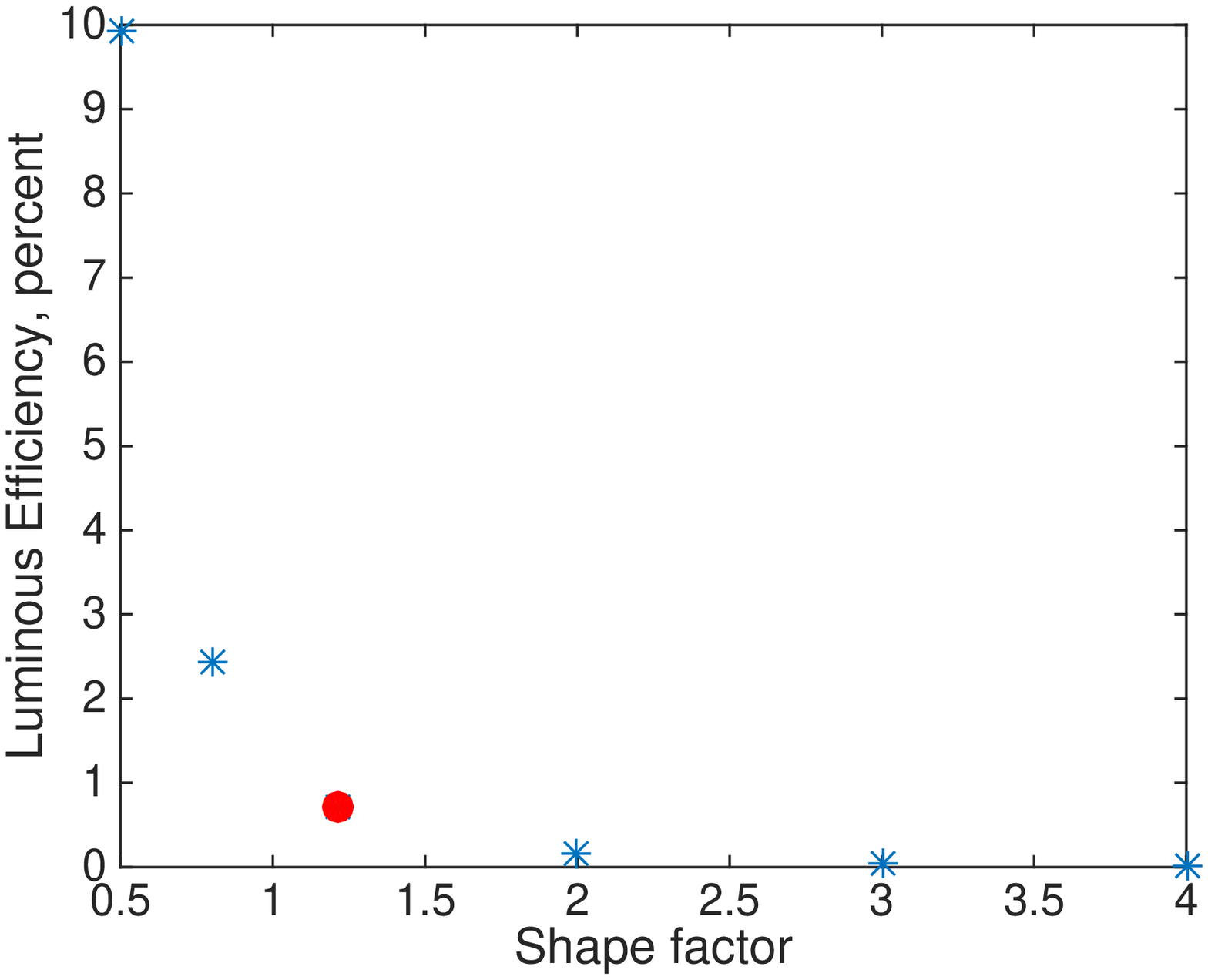}
\label{fig:shape}
\end{subfigure}
\caption{The variation of luminous efficiency with the variables assumed to be constants during ablation. The red asterisk in each figure indicates the luminous efficiency value of 0.7\% used in the standard event. The top panel shows the change in calculated luminous efficiency as a function of meteoroid density -- keeping all parameters identical and varying only the meteoroid density can cause the calculated luminous efficiency to range from 0.7\% at a density of 2000 kg/m$^3$, to 11\% at a density of 8000 kg/m$^3$. Similar changes are seen in the middle and bottom plots, for drag coefficient and shape factor. }
\label{fig:scaling}
\end{figure}

There are three variables in Equations \ref{eq:mass} and \ref{eq:tau} that are assumed to be constant with time: the drag coefficient, the shape factor, and the meteoroid density. These variables cannot be measured and values must be assumed. A representative event was simulated with the following parameters: initial speed 30 km/s; shape factor 1.21 (sphere); drag coefficient 1; meteoroid density 2000 kg/m$^3$, and a mass of $10^{-5}$ kg. Any difference between an assumed constant term and the simulated value will change the luminous efficiency by a scaling factor, and the variation and uncertainty in the calculated luminous efficiency (for a range of physically possible values) is shown in Figure \ref{fig:scaling}. \newline

\begin{figure}
\centering
\begin{subfigure}{1\textwidth}
\caption{Four days of atmospheric density data taken from the NRLMSIS E - 00 model, from 2015. The shorter red line is the simulated atmospheric density and was used in the standard event.}
\includegraphics[width=\textwidth]{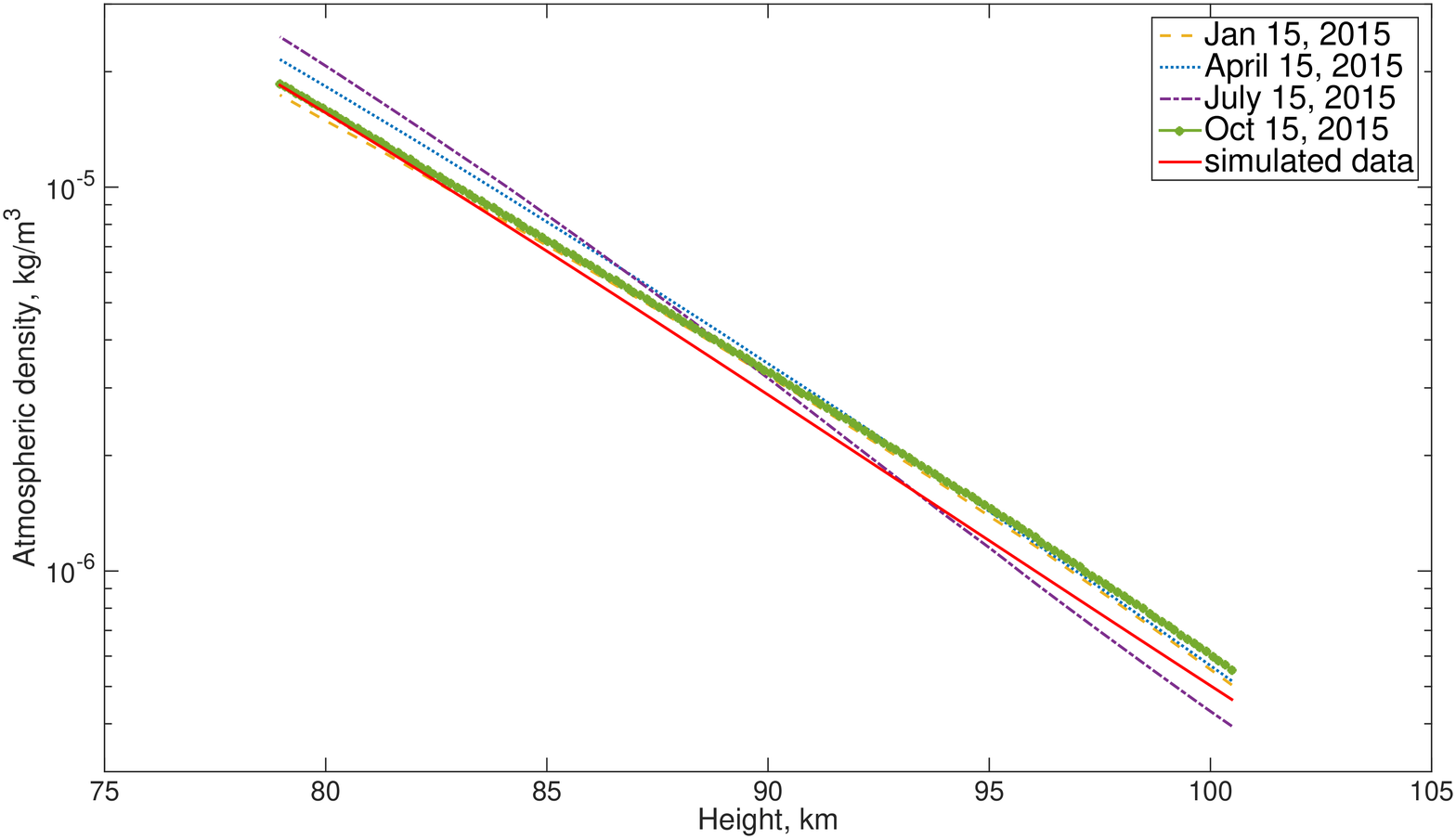}
\label{fig:atmosphere}
\end{subfigure}

\begin{subfigure}{1\textwidth}
\caption{The resulting luminous efficiencies for the four different atmospheric density models, and the model from the simulation. All parameters were kept constant except for the atmospheric density model. A luminous efficiency value of 0.7\% was used in the standard event. Note that this method is unable to reproduce the exact luminous efficiency used (0.7\% constant over time) even when using the same atmospheric density model used in the simulation.}
\includegraphics[width=\textwidth]{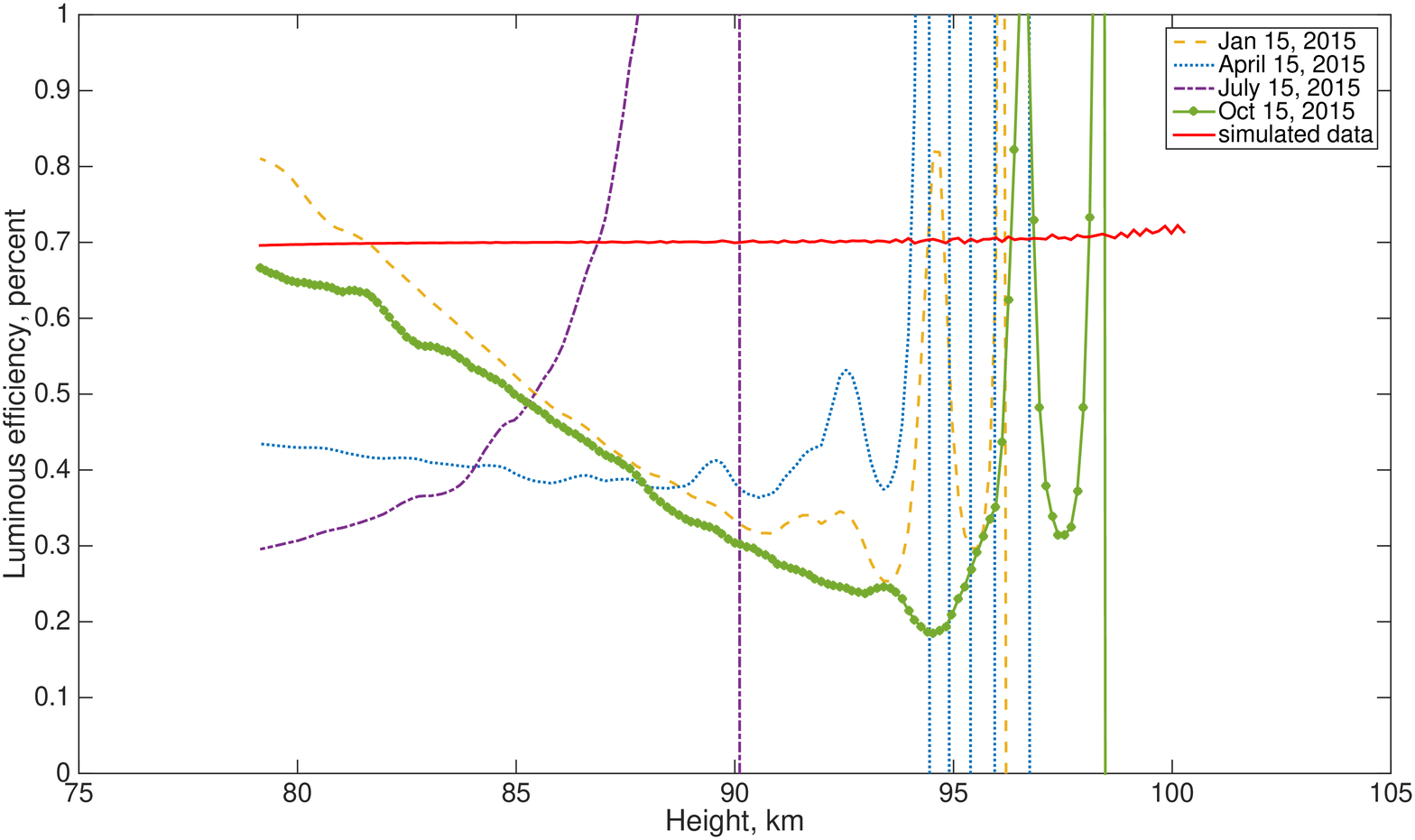}
\label{fig:atmosphere_tau}
\end{subfigure}
\caption{Atmospheric density variations over 2015 and their effect on derived luminous efficiency.}
\end{figure}

A more complicated parameter to deal with is the atmospheric density. Each of the simulated meteor events used the same atmospheric density profile, not specific to any date or location. However, with real meteor events, the atmospheric density on that day, at that time and location needs to be used. To investigate the variation in luminous efficiency due to variations in atmospheric density, four days of data (each from a different season) from the NRLMSIS E - 00 Atmosphere Model \citep{Picone2002} were compared. The four days of modelled data and the simulated atmospheric density are shown in Figure \ref{fig:atmosphere}, and the resulting luminous efficiency estimates (keeping everything the same except for the atmospheric density profile) are shown in Figure \ref{fig:atmosphere_tau}. The resulting luminous efficiency profiles show values that range from 0.2\% up to 1\%. However, not all luminous efficiency profiles have valid solutions at all heights: when the atmospheric density used in calculating the luminous efficiency is lower than the modelled atmospheric density, we end up with an unphysical situation where the meteoroid is gaining rather than losing energy as it ablates, and a singularity appears in our luminous efficiency profile. \newline

After investigating our simulated representative meteor to determine how meteoroid density, drag coefficient, shape factor, and atmospheric density model affect our calculated luminous efficiency values, the full parameter space of mass, speed, meteoroid density, zenith angle, and shape factor, was explored. Fifty meteors were simulated for each mass - speed group. The simulated meteors had a mass of $10^{-4}$, $10^{-5}$, or $10^{-6}$ kg. The speeds used were 11, 20, 30, 40, 50, 60, and 70 km/s. This meant there were 21 possible groups; however some of the low mass - low speed groups did not produce enough light that they would be detected by the CAMO optical system. This reduced the number of mass - speed groups to 18. The luminous efficiency for each meteor in this set of simulated meteors was 0.7\%, constant over time. \newline

To investigate how sensitive our results are to the chosen two-term exponential fit of the meteoroid lag, we analyzed each meteor in our mass - speed groups according to our method: the simulated position was used to determine the lag, which was fit by a two-term exponential function. This function was then numerically differentiated (i.e. finite differenced) to determine the deceleration. By using all values (drag coefficient; atmospheric density; velocity; shape factor; meteoroid density; intensity) directly from the simulation, with the exception of the determined lag, we were able to see how sensitive our derived luminous efficiency values were to our fit to the meteoroid lag. Our derived luminous efficiencies did not come out as constant values over the ablation due to the sensitivity of this method to small variations in deceleration. The mean and standard deviation for the luminous efficiency of each meteor was determined, and the average of those values in each mass - speed group are given in Table \ref{tab:mass_speed}. Fitting a two-term exponential to the lag, to find the deceleration and the luminous efficiency seems to work for most mass - speed groups. Table \ref{tab:mass_speed} shows that almost all of the mass - speed groups investigated show luminous efficiency ranges that include the true value of 0.7\%. This is not the case for high-mass, low-velocity meteors (11 km/s). In fact, for each mass group, the lowest speed that produces a luminous efficiency profile does not produce a mean luminous efficiency range that includes the value that was used in the simulation.\newline

\begin{table}
\centering
\begin{tabular}{l|*{3}{c}r}
            &  $10^{-4}$ kg & $10^{-5}$ kg & $10^{-6}$ kg\\
\hline
11 km/s&0.30 $\pm$0.11 & -&- \\
20 km/s&0.57 $\pm$0.15 & 0.51 $\pm$0.07& -\\
30 km/s&0.64 $\pm$0.18 & 0.84 $\pm$0.21& 0.50 $\pm$0.14 \\
40 km/s&0.68 $\pm$0.27 & 0.86 $\pm$0.26& 0.89 $\pm$0.33\\
50 km/s&0.60 $\pm$0.26 & 0.80 $\pm$0.28& 0.90 $\pm$0.29\\
60 km/s&0.60 $\pm$0.27 & 0.76 $\pm$0.29& 0.91 $\pm$0.32\\
70 km/s&0.58 $\pm$0.27 & 0.69 $\pm$0.28& 0.92 $\pm$0.34\\
\end{tabular}
\caption{Mean luminous efficiency value (percentage) and standard deviation for each mass - speed group of simulated meteor events. Each group initially contained 50 meteors, but some events were removed from consideration because unphysical luminous efficiency values were obtained, or due to errors in the simulated data that prevented luminous efficiency values from being determined. The luminous efficiency values used for each simulated meteor was 0.7\%.}
\label{tab:mass_speed}
\end{table}

\begin{figure}
\centering
\includegraphics[width=1.5\textwidth,center]{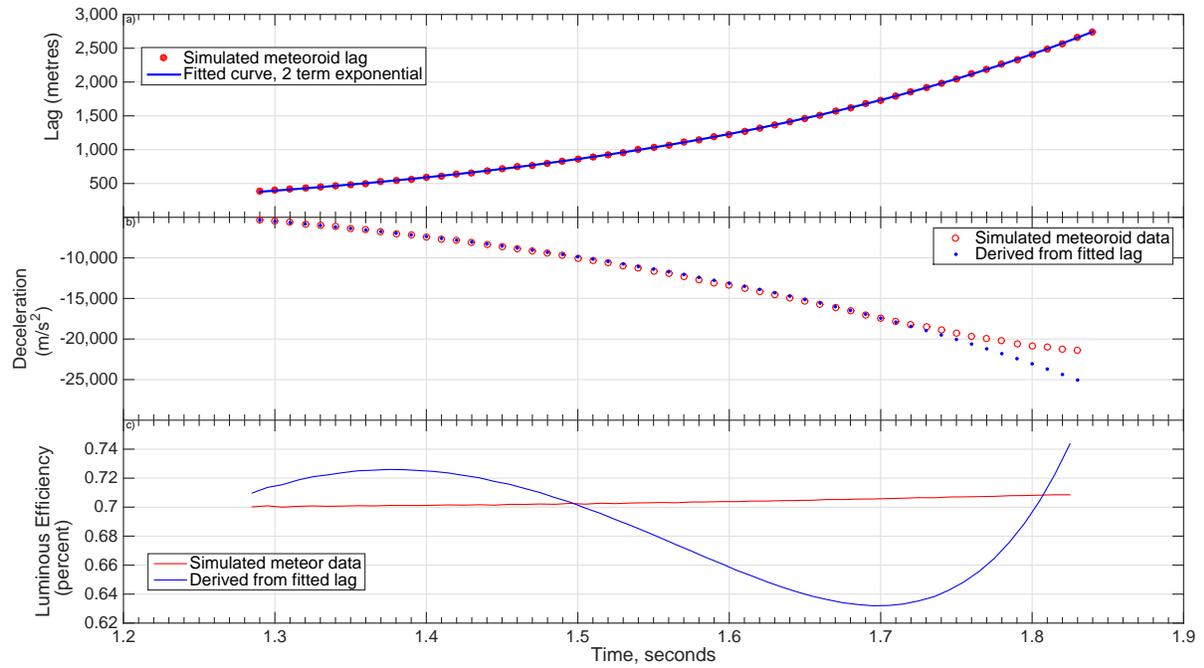}
\caption{A comparison of the output from a simulated meteoroid to fitted equivalents based on a two-term exponential fit to the simulated lag. The top panel shows the simulated lag fit with a two-term exponential function. The middle panel shows in red the deceleration produced by the simulation compared to the blue dots produced by taking the second derivative of the fit to the lag. The bottom panel shows in red, the luminous efficiency determined using only parameters output from the ablation model, while the blue line shows the luminous efficiency derived using identical parameters aside from the deceleration, which was derived from the fit shown in the top panel.}
\label{fig:typical}
\end{figure}

A comparison of the fitted lag, the corresponding deceleration, and the resulting luminous efficiency profile of a typical event are shown in Figure \ref{fig:typical}. While the simulated lag appears to be fit well by the two term exponential, the resulting deceleration from the fit deviates from the simulated deceleration. This may be due to temperature fluctuations in the ablation model. The luminous efficiency derived using only the ablation model output is unable to produce the exact luminous efficiency (constant 0.7\% over the ablation period) used in the simulation, as shown in Figure \ref{fig:shape}.  \newline

\section{Discussion}
Our method for determining the luminous efficiency uses only the luminous intensity and drag equations, while the ablation model by \citet{CampbellBrown2004} is more sophisticated. \cite{CampbellBrown2004} use the classical form of the drag equation, but their mass loss equation is not classical: they instead use the Knudsen-Langmuir formula with the Clausius-Clapeyron equation to simulate the meteoroid ablating as soon as it begins heating up. When the meteoroid becomes very hot, a spallation term is included. The third differential equation used in the \citet{CampbellBrown2004} model is the temperature equation, which describes the energy gained (through collisions with the atmospheric atoms) and lost (through radiation and evaporation of material). \newline

As seen in the previous figures, uncertainties in each of the variables of Equations \ref{eq:mass} and \ref{eq:tau} yield corresponding variances in the computed luminous efficiency. Figures \ref{fig:drag} and \ref{fig:shape} were created assuming the drag coefficient and shape factor are constant over the ablation. This is not necessarily true for real meteor events, but for simplicity, was assumed for this work, both in the modelling and analysis. If a real meteor event has a constant drag coefficient or shape factor, but an incorrect value is assumed in the analysis, the difference will be a simple scaling factor; if the value changes over the course of the flight the luminous efficiency will be off by an amount proportional to the difference in the assumed value and the average of the true value.\newline

It is obvious that variations in the atmospheric density over the course of a year (even as much as a factor of two) can change the derived luminous efficiency profile. The solid red line in Figure \ref{fig:atmosphere_tau} indicates the calculated luminous efficiency using the same atmospheric density model that was used in the simulation. A constant luminous efficiency of 0.7\% was used in the simulation, but this method is unable to exactly reproduce that: the calculated luminous efficiency is quite close to, but not exactly, a constant 0.7\%. We find that small rounding errors in the ablation model cause the small variations we see in the luminous efficiency.\newline

One of the most challenging aspects of this work is determining which functional form to fit to the lag; while more complex functional forms are able to fit the lag better (a combination of an exponential with a polynomial, for example), they do not necessarily provide a better fit to the deceleration, to which the luminous efficiency is very sensitive. Various combinations of exponential fits with polynomials (lag = a$e^{bx^2 + cx + d}$; lag = a$e^{bx}$ + cx + d; lag = a$e^{bx} + e^{cx}$ + dx + f; etc.) were tried. Much better results were obtained when the modelled deceleration was fit directly, but this approach will not work for real data. Even very precise observations from CAMO have enough noise in the measured lag that finite differencing produces wildly oscillating decelerations. A smooth fit to the lag is crucial in order to obtain a useful second derivative.\newline

We found that the luminous efficiency calculated by fitting the lag with a two-term exponential did not reproduce the model's constant 0.7\%  (see Table \ref{tab:mass_speed}). 
On average, this fitting method does return the correct luminous efficiency, except in the lowest speed groups. In particular, high mass ($10^{-4}$ kg) meteors with initial speeds of 11 km/s had a much lower mean luminous efficiency, because there was poor agreement between the simulated lag and the two-term exponential functional fits by visual inspection. Visual inspection also determined that visually good fits to the lag data may or may not produce a good match to the simulated deceleration. \newline

\section{Conclusion}
An attempt at quantifying the uncertainty in using the classical meteor ablation equations to determine the luminous efficiency of meteors has been made. Certain parameters (drag coefficient; meteoroid density; shape factor) were assumed to be constant. The wrong drag coefficient could produce errors of roughly a factor of 2; the meteoroid density can vary by a factor of 8, but is much more likely to be within a range of a factor of 2; and the shape factor may be different from a sphere, but is not very likely to be as extreme as the end values modelled here, which correspond to an oriented needle and a disk with its largest dimension oriented to the airstream. It's much more likely that the shape factor will be within a factor of 2 of a sphere, and therefore these three parameters together are each a small random effect on the luminous efficiency. The atmospheric density over the course of a year changes by a factor of 2 in the height range that meteors are detected with our optical system, and these variations cause similar factor-of-2 discrepancies in the luminous efficiency computed for simulated events. The possibility of using radar echo decay measurements to verify atmospheric density profiles at the location of the optical cameras is being investigated. Simulated meteor events were studied by examining how different functional fits to the simulated meteoroid lag and derived deceleration affected the luminous efficiency computed for each simulated meteor. A simple two-term exponential fit to the lag provides reasonable decelerations, which in turn provide an average luminous efficiency value close to what was used in the simulation. This method however, was only tested on simulated events that were free of noise. In a future work, we will test the method with noise that approximates the noise observed with the CAMO optical system, and then on actual meteor events recorded by CAMO that show single-body ablation. Measuring luminous efficiencies requires precise measurements and a thorough knowledge of the sources of uncertainty. The high-resolution CAMO tracking system will allow luminous efficiencies to be calculated much more accurately than previous observational attempts, and should be able to reveal the order of magnitude of the luminous efficiency and any trend in luminous efficiency with speed.

\section{Acknowledgements}
This work was supported by the NASA Meteoroid Environment Office [grant NNX11AB76A]. DS thanks the province of Ontario for scholarship funding. 

\newpage
\bibliography{mybibfile}

\end{document}